# 10 dB emission suppression in a structured low index medium

Soumyadeep Saha[1], Meraj E Mustafa[2, a)], Manfred Eich[2,3] and Alexander Yu. Petrov[2,3]

[1] *Waterloo Institute for Nanotechnology, University of Waterloo, Waterloo, ON, Canada.*
[2] *Institute of Optical and Electronic Materials, Hamburg University of Technology, Germany.*
3) *Institute of Photoelectrochemistry, Helmholtz-Zentrum Geesthacht, Germany.*

a) *Corresponding to: meraj.mustafa@tuhh.de*

**Abstract:** Significant suppression of radiation in 3D structured media with small refractive index 1.4-1.6, such as of glass or polymers, is a desirable feature yet to be obtained. For periodical structures this is realised at frequencies of the complete photonic band gap (CPBG), which up to now was demonstrated to open for materials with refractive index of at least 1.9. We present here a quasiperiodic 3D structure consisting of multiple overlapping gratings with a homogeneous distribution of Bragg peaks on a sphere in reciprocal space, which allows efficient suppression of emission. Recently we have presented the theoretical model, considering interaction with the neighbouring gratings only, that estimates a finite CPBG for arbitrarily small refractive indices and thus complete emission suppression in infinite structures. However, numerical simulations demonstrate a finite leakage of power from emitter not predicted by the model. Still the simulations show -10 dB suppression in 3D structures with optimised number of gratings. Astonishingly, as we show here, this limit is almost independent of the refractive index contrast. Also, the structures with a defined number of gratings show maximal suppression at certain refractive indices, losing the suppression even at higher refractive indices. The -10 dB suppression is demonstrated for refractive index contrast as low as 1.30.

## I. INTRODUCTION

Photonic crystals are periodic dielectric structures that are designed to form the energy band structure for photons, which either allows or forbids the propagation of electromagnetic waves of certain frequency ranges[1-5]. However, the limited symmetry of photonic crystals allows the opening of a complete photonic bandgap (CPBG), band gap for all directions, only at relatively high refractive index (RI) contrast[6-8]. The search for the structures with a CPBG at lowest contrast is still open. It is unclear if there is a minimal refractive index required to observe CPBG in 2D and 3D. While 2D structures have shown some decent CPBG opening at lower contrast[7-9], 3D periodic structures have not shown a CPBG opening for refractive index contrast below 1.9[6].

The low rotational symmetry of photonic crystals[10] led to the study of photonic quasicrystals for a CPBG[11-17]. In the case of quasicrystals, a periodic unit cannot be defined and thus a band gap in the band diagram is also indefinable. To discuss the CPBG in quasicrystals two approaches can be followed. A periodic approximant of the quasicrystal can be defined[9, 27] and a corresponding band diagram or local density of states can be determined by simulation of emission into aperiodic structure, which should exponentially decay with simulation volume[18, 21, 28]. A 2D quasicrystal has been reported to have strong improvement in contrast to the crystal counterpart[7,13-15]. Papers reported a single polarization CPBG in 2D structure at very low contrast[12,27]. Using the concepts similar to Ref. [27] but choosing only one ring of gratings and random phase between them we have shown that the refractive index contrast can be arbitrary small in 2D case[18]. In case of 3D quasicrystals, there were some investigation of localisation phenomena[16,17,30], however, no CPBG has been reported so far. Extending the concepts to 3D, we have developed 3D structures with an optimal distribution of Bragg peaks in reciprocal space for maximal utilisation of refractive index contrast[18]. While the theoretical calculations did predict a CPBG for any arbitrary RI contrast in both 2D and 3D structures, only the existence of CPBG in 2D structure has been confirmed by numerical simulations. For 3D structures we have observed strong but not infinite emission suppression in the structures approaching infinite size. In this paper, we extend the work on 3D structures to understand more about the obtained emission suppression. A model has been proposed to differentiate between the evanescent and propagating fields inside the quasiperiodic structure. We have evaluated structures of different refractive index contrast and number of gratings in a large simulation volume. We report such characteristics as maximal emission suppression, decay constant and suppression bandwidth.

## II. QUASICRYSTAL MODEL

The 3D quasicrystal model used in our simulations is generated as a binarized summation of sinusoidal gratings[18]. The Fourier transform of each sine function represents a pair of diametrically opposite Bragg peaks in the reciprocal space. To effectively maximize the width of the complete photonic bandgap, the Bragg peaks need to be homogenously distributed over the entire angular range[19]. This can be achieved by using known optimal solution for points distribution on a sphere with minimized maximal distance of any point on the sphere from the closest one of the points from the distribution. To preserve central symmetry known solutions with icosahedral symmetry were taken[20]. The model of the quasiperiodic refractive index distribution can be mathematically expressed as[18]

$$\Delta n(\vec{r}) = \Delta n \cdot \text{sgn}\left\{\sum_{i=1}^{N} \sin(\vec{g}_i \cdot \vec{r} + \phi_i)\right\}, \qquad (1)$$

where $\Delta n$ is the refractive index deviation from the mean value $\bar{n}$, $N$ is number of gratings used to generate the quasicrystal, $\vec{g}_i$ are the wave vectors defining the grating periods $a_i = 2\pi/g_i$ and directions and $\phi_i$ are the corresponding phases. The sign function is applied to binarize the continuous function. We took all gratings with the same period and amplitude, but random phase. The phase randomisation allows avoiding artefacts in the centre of the simulation volume and helps to derive strength of the original gratings in the binarized structure. Simulations were carried out for three different number of gratings $N = 46, 61$ and $81$ and for RI contrasts ranging from 1.2 to 1.7.

According to Bragg condition, each grating will open a bandgap along its normal direction and a bandgap at an elevated frequency in the direction deviating from the normal[18]. In our structures maximal deviation from normal direction is obtained in the direction between 3 neighboring Bragg peaks. The angle $\alpha$ between normal and maximal off-normal direction can be expressed in terms of number of gratings $N$ by approximating the Bragg peak distribution on a sphere with a hexagonal pattern. This approximation will slightly underestimate the value of $\alpha$ since coverage of the sphere with perfect and equal hexagons is not possible. The estimated and real maximal angle $\alpha$ for different number of gratings are given in TABLE I. We neglect the small deviation between estimated and real maximal value and use the estimated value in our calculations.

**TABLE I.** The estimated maximal and minimal angle α between the Bragg peaks for different grating numbers N.

| N | $\alpha_{est}$ (°) | $\alpha_{max}$ (°) |
|---|---|---|
| 46 | 13.14 | 13.68 |
| 61 | 11.41 | 11.69 |
| 81 | 9.9 | 10.19 |

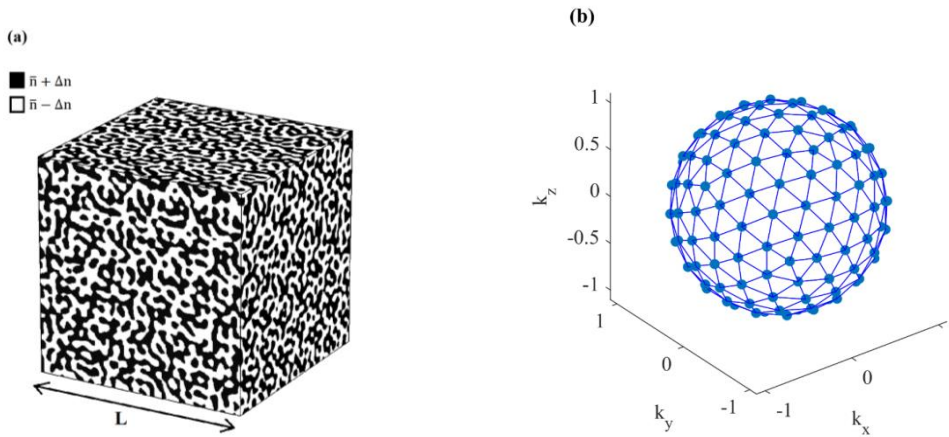

**FIG. 1.** (a) An example of a 3D quasicrystal in real space generated by overlapping 81 gratings, with a side length $L = 27.3\ a$, $a$ being the grating period. (b) The distribution of Bragg peaks in reciprocal space. The dimensions in reciprocal space are normalized by $2\pi/a$.

The quasicrystal is expected to show a CPBG if the upper edge of the PBG along the direction normal to the grating is higher than the lower edge of the PBG along the direction at angle $\alpha$. For the calculation of the lower band edge at angle $\alpha$ the effect of three neighbouring gratings is taken into account. The calculated omnidirectional PBG width can be expressed as[18]

$$\frac{\Delta\omega_{3D}}{\omega} = \left(\frac{1+\sqrt{3}}{2}\right)\sqrt{\frac{4}{\pi}\frac{\Delta n}{\bar{n}\sqrt{N}} - \frac{2\pi}{3\sqrt{3}N}}, \quad (2)$$

From this model, an optimal number of gratings exist for any refractive index contrast while a finite band gap is predicted for any finite contrast. The model neglects interaction with the gratings which are oriented further than angle $\alpha$ from the direction of observation.

### III. MODEL FOR RESIDUAL EMISSION INTO THE QUASICRYSTAL

The power emission into the quasicrystal from a small dipole at CPBG frequency is expected to decay exponentially with the increase in structure size[21]. However, as was observed from the 3D simulations[18], there is a leakage of power inside the proposed 3D quasiperiodic structure regardless of its size which limits the emission suppression. Although the mechanism of the leakage is not captured by the current model, from the observations, we propose a phenomenological description. We approximate the power emitted into the quasicrystal as

$$P(L) = P_p + P_e e^{-\gamma L/2}, \quad (3)$$

where $P_p$ is the power of propagating field, $P_e$ is the power amplitude of the evanescent field tunnelling through the structure from dipole to the outer boundary and $\gamma$ is the decay constant. The half comes from the fact that the field from dipole in the centre of the simulation volume has to tunnel through half of the structure size. Thus, this description assumes that there are two contributions to the emission from the source. Without the propagation term there is an exponential decay of emitted power with structure size. The electromagnetic field excited by the antenna is purely evanescent in this case and the emitted power corresponds to the outcoupling of the evanescent tails at the boundary of the quasiperiodic structure. The propagation term corresponds to the propagating fields which always transport energy and thus are not dependent on structure size.

### IV. SIMULATION RESULTS AND DISCUSSION

The refractive index distributions were numerically generated in MATLAB and simulated using the time domain solver of CST Microwave Studio[22]. An emitter dipole was placed exactly at the centre of the cubic structure with lateral size $L$. It should be mentioned that in a cubic structure the distance from dipole to the surface is different depending on the direction. Thus, the distance $L/2$ in Eq (3) is a minimal distance. The radiation resistance of the dipole inside the quasicrystal was obtained from the simulations. The normalised power emission spectrum $P/P_0$ of the dipole inside the structure can be obtained by calculating the normalised real part of dipole's radiation resistance $\text{Re}(Z)/\text{Re}(Z_0)$, where the normalisation was done according to the dipole's emission inside a homogenous medium of mean refractive index $\bar{n}$ [23,24]. The ratio can deviate from one at low frequencies as the dipole can be either in the low or high index material. The resulting ratio was re-normalised to 1 at lower frequencies since we are interested in the suppression with respect to the level at small frequency, or effective medium limit.

Simulations of low index 3D structures require large simulation volumes, sometimes well beyond the maximum permissible value. In our case, with a grating period $a$ of 220 nm and resolution of 20 nm, simulation can only be performed up to a maximum side length of 25 µm (*113.63a*). The choice of 220 nm period was taken to have photonic band gap in the visible, but the period and correspondingly the structure can be scaled to any wavelength. The saturation values of suppression have been determined by applying the proposed fitting model given in Eq. (2) to the simulated data. The fitted curves were determined using non-linear least square method and optimised using trust region algorithm[31]. The entire fitting procedure was carried out in logarithmic scale of the normalised emission for a balanced error calculation for small emission values.

Structures with different side lengths, up to 25 µm, were simulated for each combination of grating number $N$ and RI contrast. An example of normalised power emission spectrum for RI contrast $n_1/n_2 = 1.3$ and $N = 81$ is presented in Fig. 2. Rest of the simulated data and fitting parameters values can be found in the Supplementary Materials.

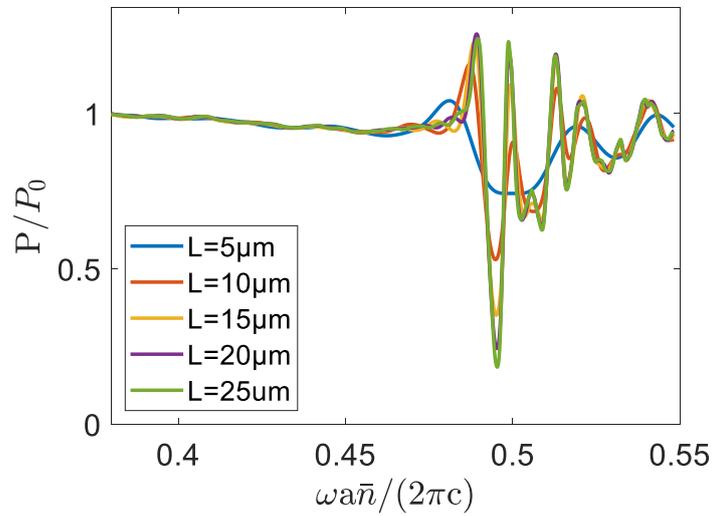

**FIG. 2.** Emission spectrum of a dipole placed inside a 3D quasiperiodic structure with RI contrast $n_2/n_1 = 1.3/1$ and icosahedral distribution of $N = 81$ underlying gratings. All spectra are divided by the dipole emission in a homogenous medium of refractive index 1.15 and then divided by the obtained ratio at normalised frequency 0.38. Thus, all normalised ratios are 1 at the left edge of the graph.

From the spectrum, a finite bandgap opening can be observed at normalised frequency of 0.513. The minima of the spectra have been plotted versus side length $L$ for $N = 81$ and for different RI contrasts in Fig 3. It can be clearly observed that the power suppression saturates at large structure size. The finite leakage into the quasiperiodic structure can be attributed to several effects not included in the theoretical estimation like polarization effect, spurious gratings introduced by binarization or interaction between non-neighbouring gratings. It can be observed that maximal suppression is achieved for intermediate refractive index and at RI contrast above 1.43 the suppression ceases again. The fact that suppression ceases at large RI contrast is also not predicted by the model. The results show that there is an optimal refractive index for maximal suppression with a given number of gratings.

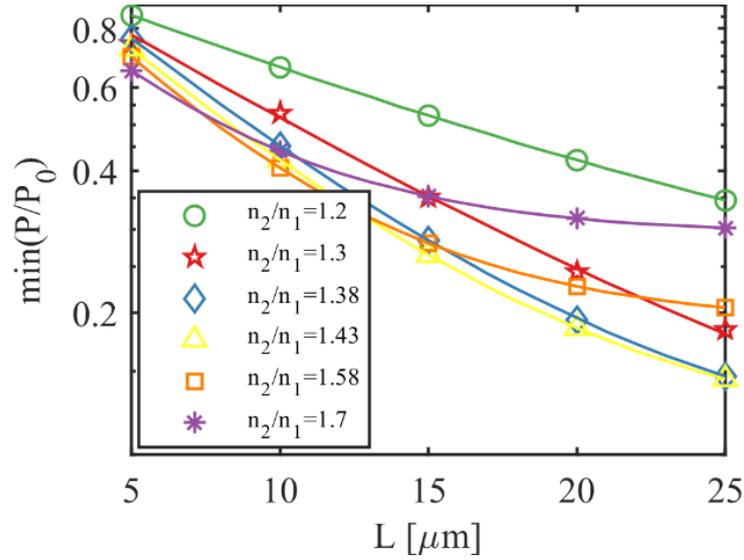

**FIG. 3.** Logarithmic plot of minima of normalised power emission over edge length $L$ for $N = 81$. The line represents fitting curve according to Eq. 3.

From the fitting parameters, the leaked power and thus maximal suppression has been calculated and plotted for different number of gratings and different RI contrasts (Fig. 4). For each number of gratings, the suppression is maximal (or in other words, the leakage is minimal) for a particular RI contrast. At higher contrast, the assumption of non-interacting gratings fails, and Bragg peak distribution does not play a decisive role any more while at lower contrast, even the expected CPBG closes due to the insufficient number of gratings. For larger number of gratings, the maximum suppression occurs at smaller RI contrasts. The maximal suppression limits to a value of approximately -10dB irrespective of different grating number. Although we could not evaluate structures with larger number of gratings and smaller refractive index from the obtained dependencies we expect that -10 dB suppression can be obtained for even smaller refractive index contrast if sufficient number of gratings is taken. This by far exceeds the suppression shown in previous works investigating 3D structures at similar RI contrasts[25,26].

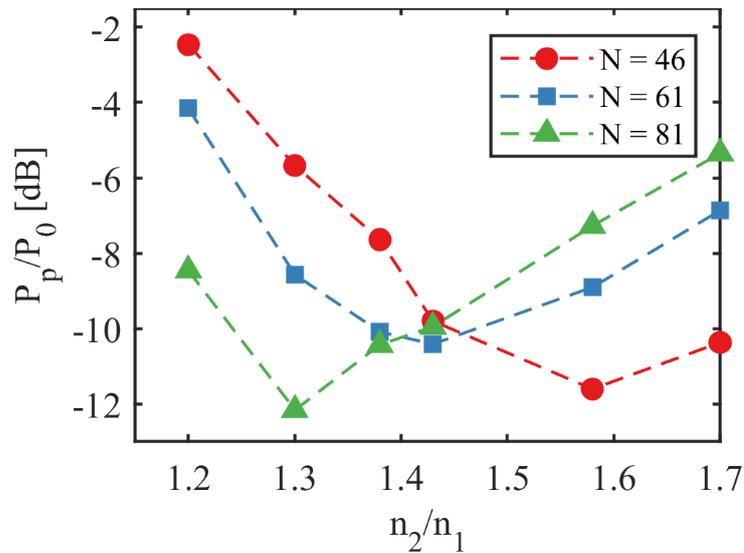

**FIG. 4.** Plot of maximal suppression (in decibel) corresponding to the normalised power of propagating fields $P_p$ observed over different RI contrast for different underlying number of gratings $N$.

The decay constant $\gamma$ has been plotted versus RI contrast in Fig. 5. Theoretically, it is expected that the decay constant should monotonically increase with RI contrast. For propagation along a one-dimensional grating, the decay constant should grow linearly with RI modulation of the grating and as the RI modulation of each partial grating increases linearly with $\Delta n$[18]. RI contrast increase from 1.2 to 1.7 corresponds to 3.5 times $\Delta n$ increase from 0.1 to 0.35. We do not observe clear linear dependence for $N = 61$ and 81 but still a growth of approximately two times. Astonishingly, for 46 gratings the decay constant is almost refractive index independent.

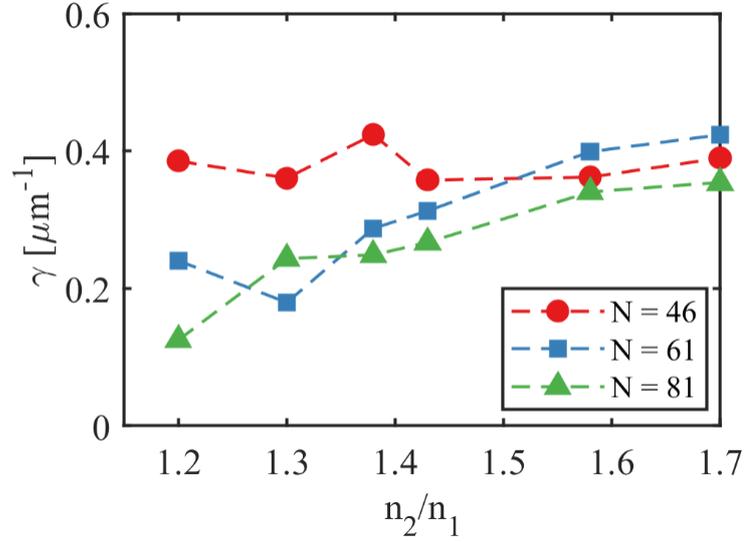

**FIG. 5.** Plot of decay constant (γ) over different RI contrast for different underlying number of gratings $N$.

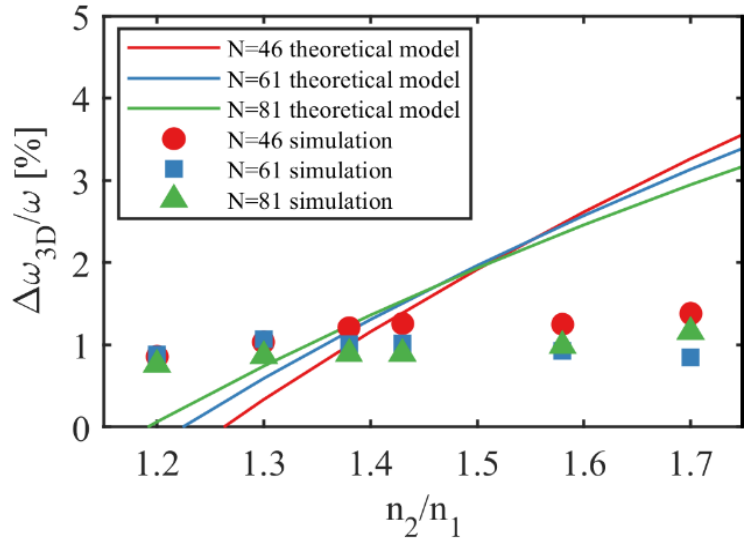

**FIG. 6.** Relative emission suppression width versus RI contrast for different number of gratings $N$. Theoretical lines are expected CPBG calculated from Eq. 2.

We also calculated the bandwidth of the emission suppression for largest structure size of 25 μm and plotted in Fig. 6. The bandwidth is defined as frequency width at the level halfway between the minimal normalised emission and 1 (see Fig. 2). Even though the theoretical calculation in Eq. 2 predicts a small yet finite bandgap for the smallest considered RI contrasts and an increase in bandgap width for larger RI contrast, the simulated data shows almost constant relative bandwidth of 1% for all N and RI contrasts. It shows that the simplified model does not describe the band gap opening correctly in the 3D structure. At the same time, it should be mentioned that the evanescent waves decay in a maximal structure of approximately 50 lattice constants or over 100 wavelengths. Thus, the minimal frequency bandwidth that can be resolved cannot be smaller than 1%, which fits to the observed bandwidths. Therefore, the obtained minimal bandwidths of emission suppression at small RI contrast are limited by the size of the structure.

## V. CONCLUSION

In conclusion, we have extended the previous study[18] of the distributed quasiperiodic structure in 3D for emission suppression at small RI contrast. There is still a finite leakage of power at the expected band gap frequency which is not captured by the theoretical model that considered overlap of directional band gaps[18]. Now a fitting description has been proposed to split the fields excited by the dipole into the evanescent and propagating part. Even though an omnidirectional photonic bandgap cannot be confirmed, we have shown numerically that the 3D structure can have -10 dB suppression of emission for RI contrast of 1.3. This by far exceeds the suppression of any previous work investigating 3D structures. We expect a similar kind of suppression for even lower RI contrast by increasing the number of gratings, however demonstrating that will require larger number of gratings in the quasiperiodic structure and larger simulation volumes. Even though the exact mechanism of the power leakage is yet to be figured out, we presented a comprehensive study. Even if the complete photonic band gap might not be possible in low refractive index 3D structures the question of the minimal leakage is still open.

**ACKNOWLEDGEMENTS**

The authors acknowledge the technical support from Dassault Systemes with their CST Studio Suite software and the funding from Deutsche Forschungsgemeinschaft (DFG) (Project number: 278744289).

**DATA AVAILABILITY**

All data is available in the manuscript or the supplementary materials. Further information on the simulation implementation and data processing is available from the corresponding author.

**COMPETING INTERESTS**

The authors declare no competing interests.

**REFERENCES**

[1]T. Maka, D. N. Chigrin, S. G. Romanov and C. M. Sotomayor Torres, "Three Dimensional Photonic Crystals in the Visible Regime," Progress In Electromagnetics Research **41**, 307-335 (2003).

[2]J. D. Joannopoulos, S. G. Johnson, J. N. Winn and R. D. Meade, *Photonic crystals: Molding the flow of light* (Princeton University Press, Princeton, ed. 2, 2008).

[3]J. D. Joannopoulos, P. R. Villeneuve and S. Fan, "Photonic crystals: putting a new twist on light," Nature **386**, 143–149 (1997).

[4]E. Yablonovitch, "Inhibited spontaneous emission in solid-state physics and electronics," Phys. Rev. Lett. **58**, 2059–2062 (1987).

[5]S. John, "Strong localization of photons in certain disordered dielectric superlattices," Phys. Rev. Lett. **58**, 2486–2489 (1987).

[6]H. Men, K. Y. K. Lee, R. M. Freund, J. Peraire and S. G. Johnson, "Robust topology optimization of three-dimensional photonic-crystal band-gap structures," Opt. Express **22**, 22632–22648 (2014).

[7]A. Cerjan and S. Fan, "Complete photonic band gaps in supercell photonic crystals," Phys. Rev. A **96**, 051802(R) (2017).

[8]W. Man, M. Megens, P. J. Steinhardt and P. M. Chaikin, "Experimental measurement of the photonic properties of icosahedral quasicrystals," Nature **436**, 993–996 (2005).

[9]P. N. Dyachenko and Y. V. Miklyaev, "Band structure calculations of 2D photonic pseudoquasicrystals obtainable by holographic lithography," Proc. SPIE 6182, 61822I (2006).

[10]G. Shang, L. Maiwald, H. Renner, D. Jalas, M. Dosta, S. Heinrich, A. Petrov and M. Eich, "Photonic glass for high contrast structural color," Sci. Rep. 8, 7804 (2018).

[11]D. Levine and P. J. Steinhardt, "Quasicrystals: A new class of ordered structures," Phys. Rev. Lett. 53, 2477–2480 (1984).

[12]Z. V. Vardeny, A. Nahata and A. Agrawal, "Optics of photonic quasicrystals," Nat. Photonics 7, 177–187 (2013).

[13]W. Steurer and D. Sutter-Widmer, "Photonic and phononic quasicrystals," J. Phys. D: Appl. Phys. 40, R229-R247 (2007).

[14]M. E. Zoorob, M. D. B. Charlton, G. J. Parker, J. J. Baumberg and M. C. Netti, "Complete photonic bandgaps in 12-fold symmetric quasicrystals," Nature 404, 740–743 (2000).

[15]X. Zhang, Z.-Q. Zhang and C. T. Chan, "Absolute photonic band gaps in 12-fold symmetric photonic quasicrystals," Phys. Rev. B 63, 81105 (2001).

[16]L. Dal Negro and S. V. Boriskina, "Deterministic aperiodic nanostructures for photonics and plasmonics applications," Laser Photonics Rev. 6, 178–218 (2012).

[17]S.-Y. Jeon, H. Kwon and K. Hur, "Intrinsic photonic wave localization in a three-dimensional icosahedral quasicrystal," Nat. Phys. 13, 363–368 (2017).

[18]Lukas Maiwald, Timo Sommer, Mikhail S. Sidorenko, Ruslan R. Yafyasov, Meraj E. Mustafa, Marvin Schulz, Mikhail V. Rybin, Manfred Eich, and Alexander Yu. Petrov, "Control over Light Emission in Low-Refractive-Index Artificial Materials Inspired by Reciprocal Design," Adv. Opt. Mat. 10, 2100785 (2021).

[19]W. Man, M. Florescu, K. Matsuyama, P. Yadak, G. Nahal, S. Hashemizad, E. Williamson, P. Steinhardt, S. Torquato, and P. Chaikin, "Photonic band gap in isotropic hyperuniform disordered solids with low dielectric contrast," Opt. Express 21, 19972-19981 (2013)

[20]R. H. Hardin, N. J. A. Sloane and W. D. Smith, "Tables of spherical codes with icosahedral symmetry," published electronically at http://neilsloane.com/icosahedral.codes.

[21]A. Della Villa, S. Enoch, G. Tayeb, V. Pierro, V. Galdi and F. Capolino, "Band gap formation and multiple scattering in photonic quasicrystals with a Penrose-type lattice," Phys. Rev. Lett. 94, 183903 (2005).

[22]More information available at https://www.3ds.com/products-services/simulia/products/cststudio-suite/.

[23]A. E. Krasnok, A. P. Slobozhanyuk, C. R. Simovski, S. A. Tretyakov, A. N. Poddubny, A. E. Miroshnichenko, Y. S. Kivshar and P. A. Belov, "An antenna model for the Purcell effect," Sci. Rep. 5, 12956 (2015).

[24]K. M. Schulz, D. Jalas, A. Y. Petrov and M. Eich, "Reciprocity approach for calculating the Purcell effect for emission into an open optical system," Opt. Express 26, 19247–19258 (2018).

[25]M. J. Ventura and M. Gu, "Engineering spontaneous emission in a quantum-dot-doped polymer nanocomposite with three-dimensional photonic crystals," Adv. Mater. 20, 1329–1332 (2008).

[26]H. Yin, B. Dong, X. Liu, T. Zhan, L. Shi, J. Zi and E. Yablonovitch, "Amorphous diamond-structured photonic crystal in the feather barbs of the scarlet macaw," Proc. Natl. Acad. Sci. U.S.A. 109, 10798–10801 (2012).

[27]M. C. Rechtsman, H. C. Jeong, P. M. Chaikin, S. Torquato and P. J. Steinhardt, "Optimized structures for photonic quasicrystals," Phys. Rev. Lett. 101, 073902 (2008).

[28]C. Rockstuhl and F. Lederer, "Suppression of the local density of states in a medium made of randomly arranged dielectric spheres," Phys. Rev. B. 79, 132202 (2009).

[29]P. L. Hagelstein and D. R. Denison, "Nearly isotropic photonic bandgap structures in two dimensions," Opt. Lett. 24, 708-710 (1999).

[30]A. D. Sinelnik, I. I. Shishkin, X. Yu, K. B. Samusev, P. A. Belov, M. F. Limonov, P. Ginzburg and M. V. Rybin, "Experimental Observation of Intrinsic Light Localization in Photonic Icosahedral Quasicrystals," Adv. Opt. Mat. 8, 2001170 (2020).

[31]More information available at https://www.mathworks.com/help/optim/ug/equation-solving-algorithms.html.

1) Waterloo Institute for Nanotechnology, University of Waterloo, Waterloo, ON, Canada.
2) Institute of Optical and Electronic Materials, Hamburg University of Technology, Germany.
3) Institute of Photoelectrochemistry, Helmholtz-Zentrum Geesthacht, Germany.

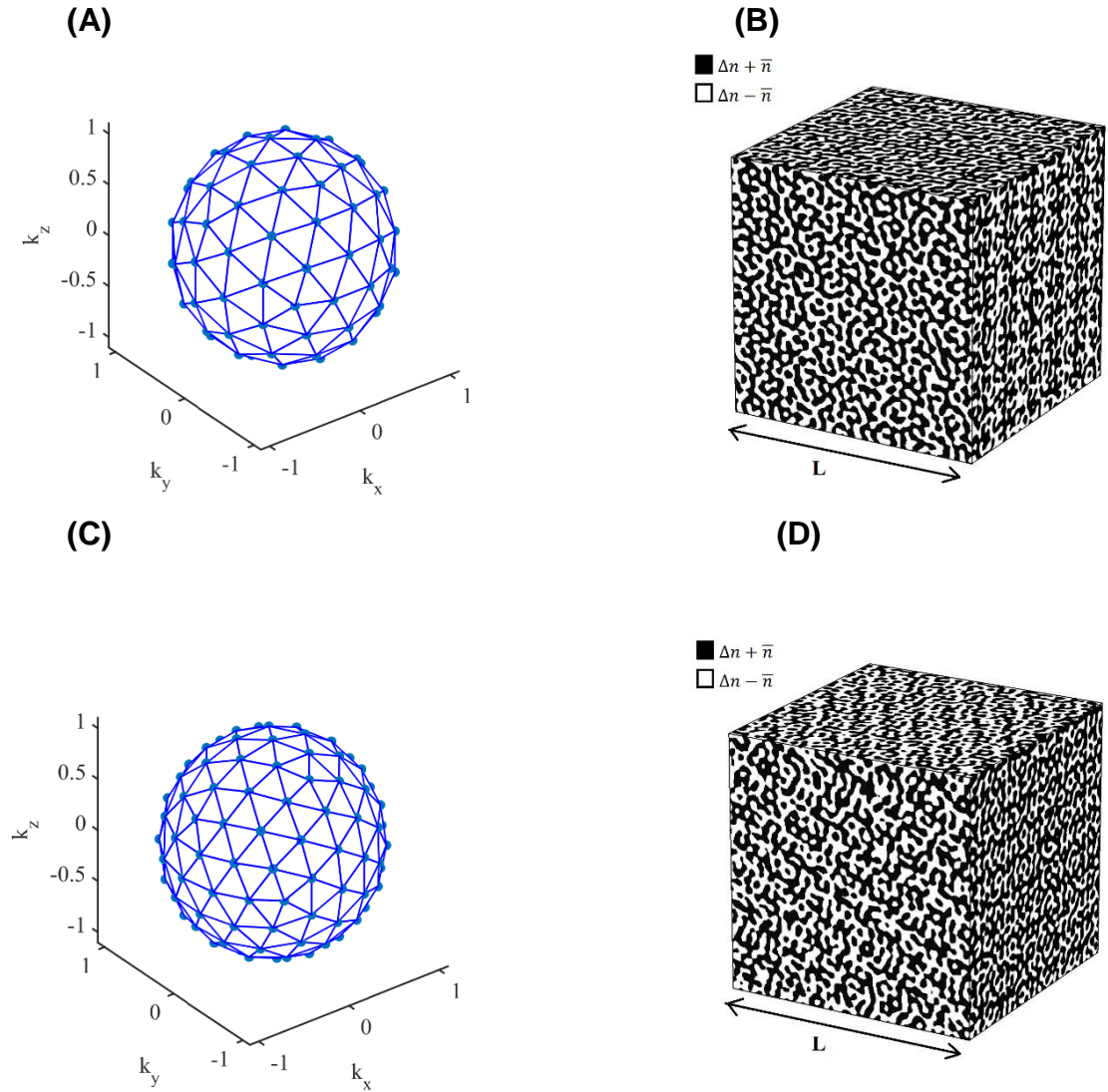

**FIG. S1.** The distribution of Bragg peaks in reciprocal space for (A) $N = 46$ and (C) $N = 61$. The dimensions in reciprocal space are normalized by $2\pi/a$, $a$ being the grating period. 3D quasicrystal in real space generated by overlapping (B) 46 gratings and (D) 61 gratings, each having a side length $= 27.3\ a$.

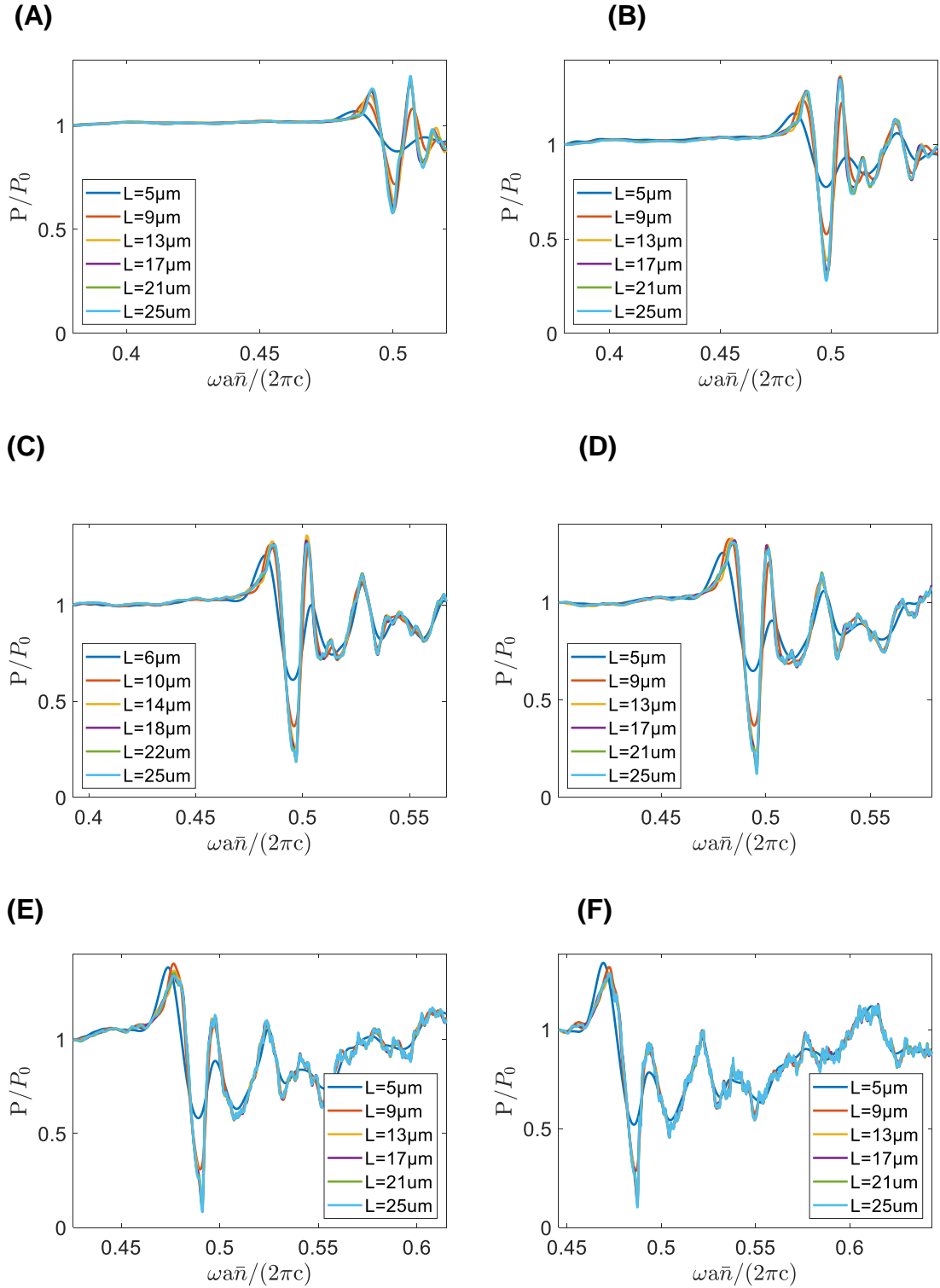

**FIG. S2.** Emission spectrums of a dipole placed inside a 3D quasiperiodic structure with icosahedral distribution of $N = 46$ underlying gratings and RI contrast of (A) $n_2/n_1 = 1.2/1$, (B) $n_2/n_1 = 1.3/1$, (C) $n_2/n_1 = 1.38/1$, (D) $n_2/n_1 = 1.43/1$, (E) $n_2/n_1 = 1.58/1$ and (F) $n_2/n_1 = 1.7/1$. All spectra are divided by the dipole emission in a homogenous medium of respective mean refractive indices and then divided by the obtained ratio at normalised frequency 0.38. Thus, all normalised ratios are 1 at the left edge of the graph.

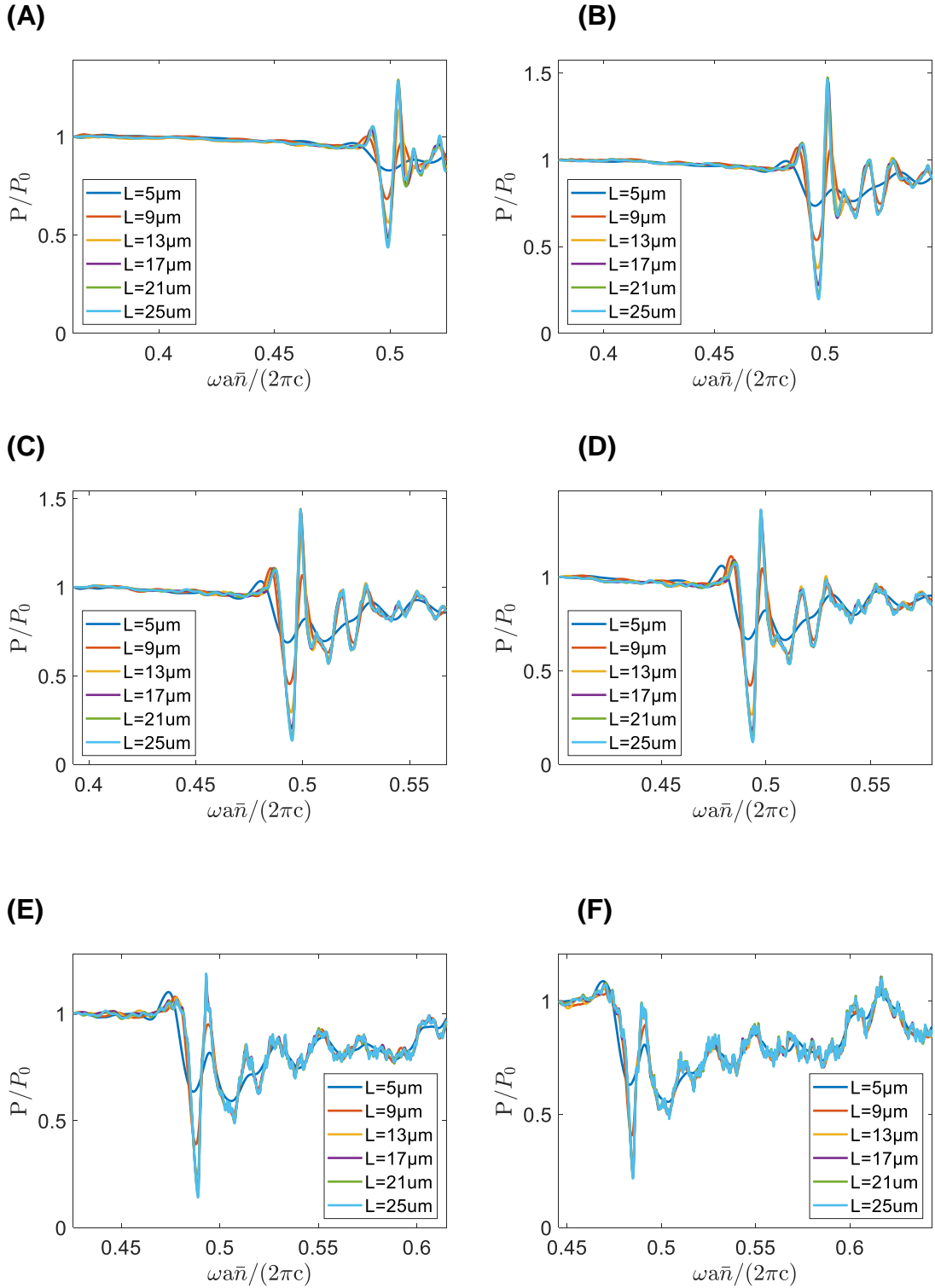

**FIG. S3.** Emission spectrums of a dipole placed inside a 3D quasiperiodic structure with icosahedral distribution of $N = 61$ underlying gratings and RI contrast of (A) $n_2/n_1 = 1.2/1$, (B) $n_2/n_1 = 1.3/1$, (C) $n_2/n_1 = 1.38/1$, (D) $n_2/n_1 = 1.43/1$, (E) $n_2/n_1 = 1.58/1$ and (F) $n_2/n_1 = 1.7/1$,. All spectra are divided by the dipole emission in a homogenous medium of respective mean refractive indices and then divided by the obtained ratio at normalised frequency 0.38. Thus, all normalised ratios are 1 at the left edge of the graph.

12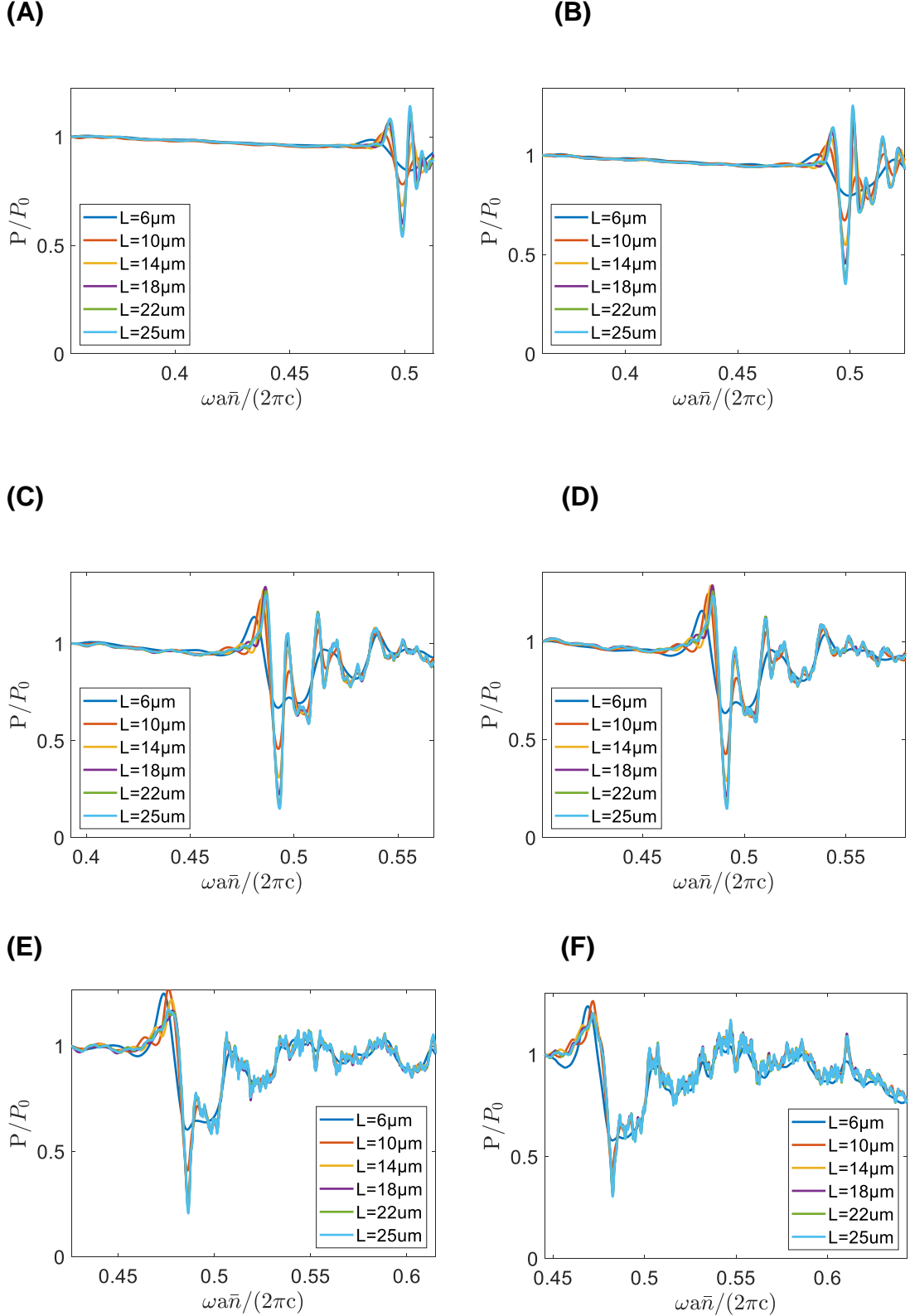

**FIG. S4.** Emission spectrums of a dipole placed inside a 3D quasiperiodic structure with icosahedral distribution of $N = 81$ underlying gratings and RI contrast of (A) $n_2/n_1 = 1.15/1$, (B) $n_2/n_1 = 1.2/1$, (C) $n_2/n_1 = 1.38/1$, (D) $n_2/n_1 = 1.43/1$, (E) $n_2/n_1 = 1.58/1$ and (F) $n_2/n_1 = 1.7/1$,. All spectra are divided by the dipole emission in a homogenous medium of respective mean refractive indices and then divided by the obtained ratio at normalised frequency 0.38. Thus, all normalised ratios are 1 at the left edge of the graph.

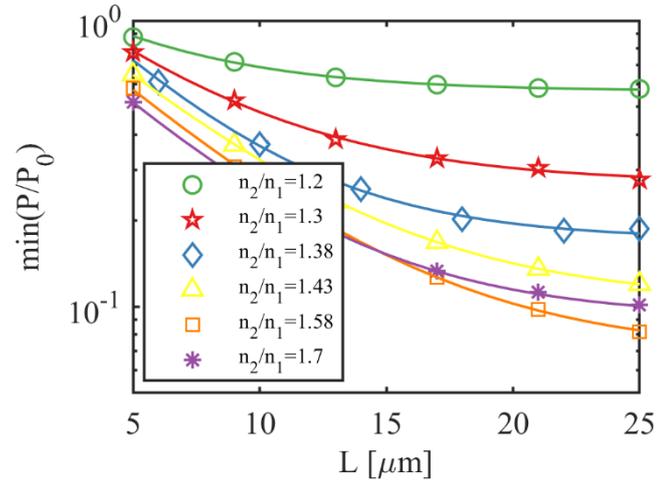

**FIG. S5.** Logarithmic plot of minima of normalised power emission over edge length $L$ for $N = 46$. The line represents fitting curve according to Eq. 3.

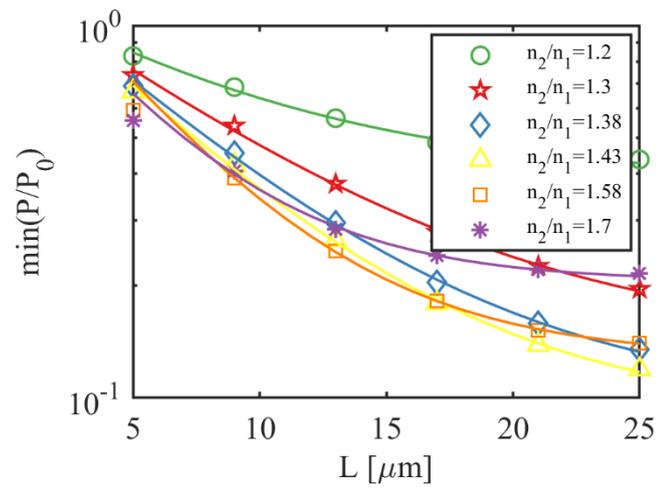

**FIG. S6.** Logarithmic plot of minima of normalised power emission over edge length $L$ for $N = 61$. The line represents fitting curve according to Eq. 3